\documentclass[runningheads]{llncs}
\usepackage[T1]{fontenc}
\usepackage{graphicx}
\usepackage{multirow}
\usepackage{amsmath}
\usepackage{url}
\usepackage{hyperref}
\usepackage{orcidlink}
\usepackage{listings}
\usepackage{subfig}
\begin{document}
\title{A Unified Ontology for Scalable Knowledge Graph–Driven Operational Data Analytics in High-Performance Computing Systems}%: Integrating M100 and F-DATA}
\titlerunning{A Unified Operational Data Analytics Ontology}

% \author{Junaid Ahmed Khan\inst{1}\orcidID{0009-0007-6131-7473} \and
% Andrea Bartolini\inst{2}\orcidID{0000-0002-1148-2450}} %\and
% %Third Author\inst{3}\orcidID{2222--3333-4444-5555}}
% %
% \authorrunning{Khan et al.}
% \institute{DEI Department, University of Bologna, Italy  \\
% \email{junaidahmed.khan@unibo.it}\\
% \and
% DEI Department, University of Bologna, Italy \\
% \email{a.bartolini@unibo.it}}

\author{
  Junaid Ahmed Khan\inst{1}\orcidlink{0009-0007-6131-7473} \and
  Andrea Bartolini\inst{1}\orcidlink{0000-0002-1148-2450}
}

\authorrunning{J. A. Khan and A. Bartolini}

\institute{
  DEI Department,\\
  University of Bologna, Italy\\
  \email{junaidahmed.khan@unibo.it} \\
  \email{a.bartolini@unibo.it}
}
\maketitle              % typeset the header of the contribution
\begin{abstract}
% The abstract should briefly summarize the contents of the paper in

Modern high-performance computing (HPC) systems generate massive volumes of heterogeneous telemetry data from millions of sensors monitoring compute, memory, power, cooling, and storage subsystems. As HPC infrastructures scale to support increasingly complex workloads—including generative AI—the need for efficient, reliable, and interoperable telemetry analysis becomes critical. Operational Data Analytics (ODA) has emerged to address these demands; however, the reliance on schema-less storage solutions limits data accessibility and semantic integration. Ontologies and knowledge graphs (KG) provide an effective way to enable efficient and expressive data querying by capturing domain semantics, but they face challenges such as significant storage overhead and the limited applicability of existing ontologies, which are often tailored to specific HPC systems only. In this paper, we present the first unified ontology for ODA in HPC systems, designed to enable semantic interoperability across heterogeneous data centers. Our ontology models telemetry data from the two largest publicly available ODA datasets—M100 (Cineca, Italy) and F-DATA (Fugaku, Japan)—within a single data model. The ontology is validated through 36 competency questions reflecting real-world stakeholder requirements, and we introduce modeling optimizations that reduce knowledge graph (KG) storage overhead by up to 38.84\% compared to a previous approach, with an additional 26.82\% reduction depending on the desired deployment configuration. This work paves the way for scalable ODA KGs and supports not only analysis within individual systems, but also cross-system analysis across heterogeneous HPC systems.

%diagnostics. %This work makes it easier to access and analyze telemetry data across diverse HPC systems, supporting more efficient monitoring, diagnostics, and decision-making.

\keywords{High-performance computing (HPC) systems  \and Operational data analytics (ODA) \and Knowledge graph (KG).}
\end{abstract}

\section{Introduction}
\label{sec:introduction}

Modern high-performance computing (HPC) systems generate massive volumes of telemetry data. Equipped with millions of sensors across critical subsystems—including compute nodes, memory, interconnects, power, cooling, and storage—these systems monitor diverse physical and performance parameters such as temperature, voltage, fan speed, and power consumption. This data is heterogeneous, high-frequency, and voluminous, creating significant challenges for system monitoring, diagnostics, and maintenance. These difficulties are compounded by rising computational demands driven by computationally intensive workloads such as machine learning and generative AI. As supercomputers grow in scale, complexity, and operational cost, ensuring their efficiency, reliability, and sustainability becomes increasingly difficult. To address these issues, Operational Data Analytics (ODA) has emerged as a key approach to address these issues by enabling telemetry data collection, storage, and analysis to support system optimization \cite{ODA}. Recent research has focused on building high-level services on top of ODA platforms—such as dashboards, digital twins, and machine learning pipelines for anomaly detection and predictive maintenance \cite{ExaDigit_SC,netti2021conceptual,graafe}. Despite these advances, access to underlying telemetry remains difficult for system administrators, facility operators, engineers, and end users. A key challenge is the reliance on schema-less NoSQL databases as the default storage solution for telemetry data \cite{ODA}. While these databases are effective for ingesting heterogeneous data, they lack predefined schemas, which complicates the writing of complex queries and often necessitates manual effort to establish connections between related data sources.

Ontologies and knowledge graphs (KGs) offer a solution to these limitations. KGs are built on explicitly defined ontologies that model data structure and domain semantics, enabling clear representation of relationships among HPC components—such as jobs executed on compute nodes. This semantic layer supports expressive querying using SPARQL query language, allowing users to access telemetry data without in-depth knowledge of underlying formats \cite{khanExaQueryProvingData2024}. Prior work on semantic modeling of cloud infrastructure \cite{imam2016applicationontologiescloudcomputing,CASTANE2018373,ont_ict_infra} primarily focuses on static descriptions of hardware, software, and services and lacks grounding in telemetry data, which is essential for ODA in HPC\cite{ODA}. Existing ODA ontologies \cite{khanExaQueryProvingData2024,xiaoyu_alexandru_graphsys24} are tailored to specific HPC environments, limiting their reuse and interoperability. Furthermore, HPC telemetry consists of high-frequency time-series data, and when encoded in RDF triples according to the ontology in \cite{khanExaQueryProvingData2024}, it incurs significant storage overhead (approx 745× compared to NoSQL) \cite{khan2025operationaldataanalyticschatbots}. These challenges highlight the need for a unified, storage-efficient ontology capable of generalizing across HPC environments, supporting consistent queries over heterogeneous ODA telemetry, and enabling cross-system interoperability.

In this paper, we present the first unified ontology for ODA in HPC systems, addressing key limitations of previous approaches. In the open-space of data center telemetry, only two publicly available ODA datasets exist from systems ranked in the top 10 of the TOP500 (\url{https://top500.org/lists/top500/2024/11/}) list of the world’s most powerful supercomputers: the M100 dataset (49 TB uncompressed) from the Marconi100 system at CINECA (\url{https://www.cineca.it/it}) in Italy \cite{m100nature}, which provides the largest available ODA telemetry and is the predecessor of the Leonardo system currently in the top 10, and the F-DATA dataset from the Fugaku supercomputer developed by the RIKEN Center for Computational Science (\url{https://en.wikipedia.org/wiki/Fugaku_(supercomputer)}) in Japan \cite{antici2024fdata}. Therefore, our proposed ontology focuses on these two heterogeneous HPC environments, aiming to model them within a single, adaptable semantic framework. In addition to unifying the representation of ODA telemetry data, the ontology introduces modeling optimizations that reduce KG storage overhead by 38.84\% compared to the previous state-of-the-art \cite{khanExaQueryProvingData2024}, with an additional 26.82\% reduction achieved by using blank nodes (i.e., no separate URIs) for sensor reading nodes, depending on the deployment configuration. We validate the proposed Unified ODA Ontology using a set of 36 competency questions designed to reflect real-world stakeholder needs. This work presents the first unified ontology for ODA in HPC systems, enabling semantic interoperability across heterogeneous data centers and laying the groundwork for scalable KGs in ODA.

\section{Related Works}
\label{sec:related_works}

In the domain of data centers and HPC, several works have proposed logical data models. For example, in \cite{ont_ict_infra}, the authors introduce an ontology that models key cloud infrastructure elements and their relationships using data from configuration and IT service management systems. Similarly, \cite{noriaO_ont} focuses on event tracking and management in cloud environments through an ontology-based approach. Although these ontologies describe data center components and their interconnections, they are primarily component-driven and do not address the modeling of operational data, which is the main focus of our work. The ontology in \cite{CASTANE2018373} integrates HPC and cloud infrastructure but also concentrates mainly on HPC–cloud relationships and does not capture ODA telemetry data.

More recent efforts have begun to address ontologies for ODA directly. For instance, \cite{xiaoyu_alexandru_graphsys24} presents an ODA ontology developed for the SURF supercomputing facility, while \cite{khanExaQueryProvingData2024} proposes an ODA ontology for the Marconi100 system. However, both are tightly coupled to specific sites and lack the generalization needed to model operational data across multiple HPC centers. Our work builds upon \cite{khanExaQueryProvingData2024} as it already models one of our target systems (Marconi100). We extend this ontology with additional concepts and relationships to make it adaptable for heterogeneous ODA data from multiple HPC sites and address storage overhead challenges inherent to RDF representations.

The remainder of the paper is organized as follows: Section~\ref{sec:ontology} presents the proposed Unified ODA Ontology; Section~\ref{sec:experimental_results} details the experimental evaluation; Section~\ref{sec:discussion} discusses the obtained results; and Section~\ref{sec:conclusion} concludes the paper. The proposed ontology and its resources are available on the project’s Git repository\footnote{\url{https://gitlab.com/ecs-lab/unified-oda-ontology}}.

\section{Unified Operational Data Analytics (ODA) Ontology}
\label{sec:ontology}

\begin{figure*}[ht]
    \centering
    \includegraphics[width=\linewidth]{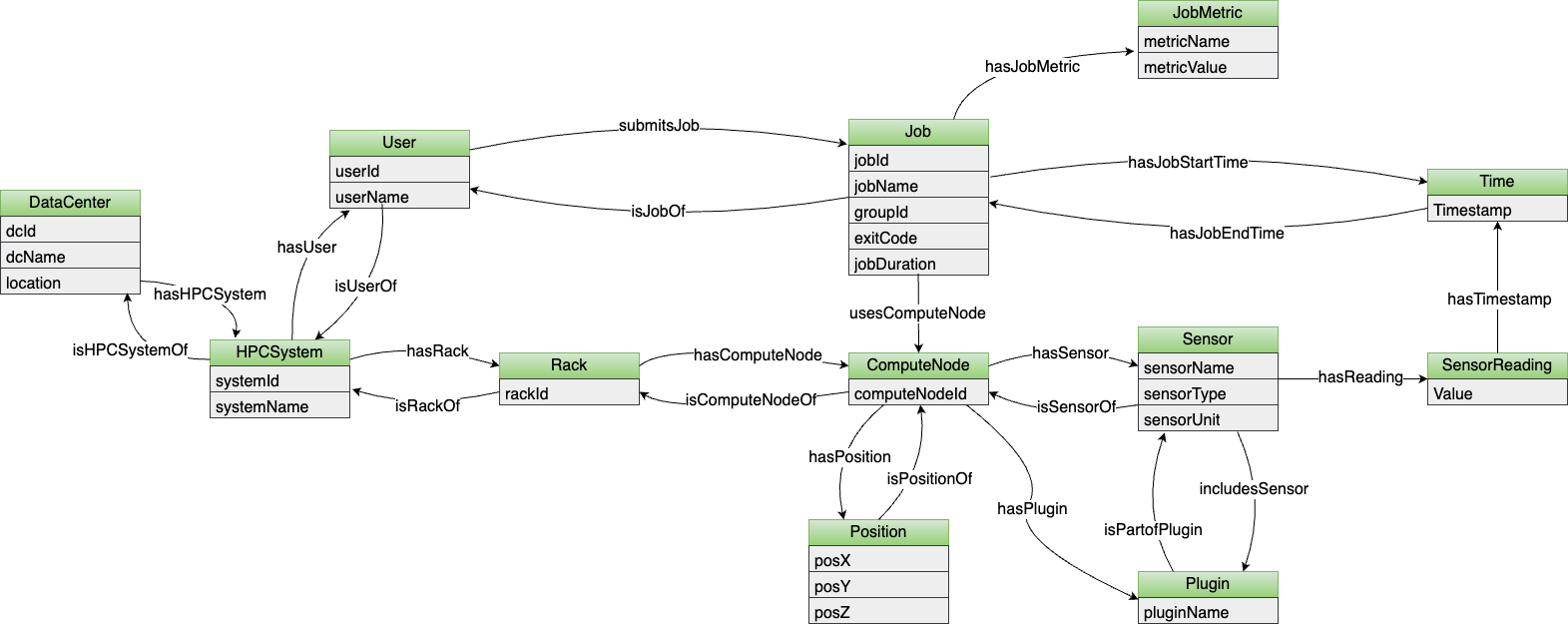}
    \caption{Unified ODA Ontology with its classes and properties.}
    \label{fig:oda_ontology_v2}
\end{figure*}

% Figure~\ref{fig:oda_ontology_v2} presents the proposed unified ontology for the HPC systems, and the Table~\ref{tab:ontology-metrics} provides a quantitative overview of the structural composition of the proposed ontology. The ontology consists of a total of 161 axioms, out of which 102 are logical axioms that define the semantics of the model—such as property characteristics, domains, ranges, and inverse relationships—while the remaining 59 are declaration axioms that introduce and declare the named entities used within the ontology. The schema includes 12 named classes, representing core conceptual entities such as jobs, compute nodes, racks, and sensors. In addition, the ontology defines 23 object properties to model inter-class relationships (e.g., rack has compute node, jobs executed on the compute node, etc), and 24 data properties that bind individuals to literal values—these include attributes such as numeric metrics, unique identifiers, and timestamped events.

Figure~\ref{fig:oda_ontology_v2} presents the proposed Unified ODA Ontology, comprising 164 axioms: 104 logical axioms defining semantics (e.g., domains, ranges, characteristics, inverses) and 60 declarations introducing named entities. The ontology defines 12 classes representing core HPC concepts—jobs, compute nodes, racks, sensors, users—and includes 23 object properties for inter-class relations (e.g., job-to-node, rack-to-node) and 25 data properties linking individuals to literals (e.g., timestamps, metrics). Designed for interoperability across heterogeneous HPC systems, the schema supports multiple facilities via \texttt{DataCenter} and \texttt{HPCSystem}, captures workloads through \texttt{User}, \texttt{Job}, and \texttt{JobMetric}, and models infrastructure layout with \texttt{Rack}, \texttt{ComputeNode}, and \texttt{Position}. Monitoring data is represented via \texttt{Sensor} and \texttt{SensorReading}, while temporal dynamics are abstracted with the \texttt{Time} class, enabling representation of time-dependent events and relationships. The \texttt{Plugin} class models software components involved in monitoring and analysis.

\subsection{Ontology design process}
\label{sec:ontology_design_process}

The design of our ontology was driven by a set of 36 competency questions (see Table~\ref{tab:competency_questions}) formulated to reflect the analytical needs of end-users—namely facility managers and data center engineers—seeking insights from ODA telemetry. These competency questions span six categories: (1) \textbf{System \& Infrastructure Topology}, which addresses the identification and spatial configuration of hardware components in an HPC system; (2) \textbf{Sensor \& Monitoring}, focusing on temporal sensor data such as temperature, CPU utilization, and power metrics; (3) \textbf{Job Metrics \& Execution Analysis}, which captures resource consumption and job performance characteristics; (4) \textbf{User \& Group Activity}, encompassing user behavior and job submission patterns; (5) \textbf{Scheduling \& Queue Analysis}, evaluating the efficiency of job queuing and scheduling patterns; and (6) \textbf{Cross-System \& Comparative Analytics}, enabling performance and efficiency comparisons across different HPC systems.

We extended the ODA ontology introduced in \cite{khanExaQueryProvingData2024}, originally developed for the M100 system at Cineca \cite{m100nature}. While effective for representing telemetry from M100, it lacked the abstraction needed to generalize across systems. To support cross-system competency questions, we introduced two new top-level classes: \texttt{DataCenter} and \texttt{HPCSystem} in the proposed ontology. These additions provide a generalized schema to distinguish and integrate data originating from different sites. Additionally, with a dedicated category of competency questions focused on user-specific analysis, we identified the need to explicitly represent users within the ontology. Consequently, we introduced the \texttt{User} class to capture user-related metadata. Furthermore, F-DATA\cite{antici2024fdata} dataset is fundamentally job-centric, where time-series telemetry is recorded per job rather than per sensor. These metrics include performance counters such as execution cycles, memory read/write volumes, as well as derived quantities like job-specific power and energy consumption. In contrast, the ODA ontology in \cite{khanExaQueryProvingData2024}, includes a \texttt{Sensor} class, that links performance data with jobs indirectly through the compute node entity where the job executes. This model proved insufficient for accurately representing F-DATA’s per-job metrics, which are not inherently tied to individual physical sensors. To resolve this limitation, we introduced a dedicated \texttt{JobMetric} class in the proposed ontology. This addition enables the direct representation of job-specific performance data and facilitates semantic alignment with both Fugaku's job-centric metrics \cite{antici2024fdata} and M100’s job table attributes \cite{m100nature}. Moreover, it establishes a clear distinction between sensor-based telemetry and job-level metrics.

%\vspace{-10pt}

To address significant scalability challenges in the KG constructed using the ODA ontology in \cite{khanExaQueryProvingData2024}—driven by the excessive number of RDF triples per sensor reading and resulting in over 700× the storage size of equivalent NoSQL formats—we introduced several schema-level optimizations. At a 20s sampling rate, one sensor alone produces 4,320 readings per day, and each reading instantiated six RDF triples: one representing the sensor reading node itself, followed by separate triples for the reading value, timestamp, and unit, as well as links to the corresponding Sensor entity and the source DataRecord from which the reading is part of in the NoSQL DB. For the 104 sensors managed by the IPMI plugin on a single compute node of the M100 system \cite{m100nature}, this schema results in over 2.6 million RDF triples generated per day. To reduce this overhead, we revised the ontology by (i) eliminating the DataRecord class, which was redundant with respect to our competency questions, and (ii) relocating the unit specification from each \texttt{SensorReading} to the \texttt{Sensor} class itself—thereby avoiding repetitive triples. These optimizations achieved a 33.3\% reduction in the total number of sensor-related triples. Additionally, we recognized another inefficiency related to timestamp representation. In \cite{khanExaQueryProvingData2024}, the ontology used ISO 8601 formatted strings (25 characters), which contributed to KG bloat. We introduced a \texttt{Time} class to centralize and reuse temporal values and adopted Unix timestamp encoding (10 characters), reducing storage demands further while preserving temporal granularity. Each sensor reading now links to a single instance of the Time class via its Unix timestamp, enabling more compact and efficient KG construction.

\subsection{Classes and Properties}
\label{sec:classes_properties}

The ontology defines 12 classes, representing core concepts within the data center infrastructure. Table~\ref{tab:classes} provides an overview of these classes with a brief description.

\vspace{-25pt}
\begin{table}[ht]
\centering
\caption{Classes and their descriptions in the proposed Unified ODA Ontology.}
\label{tab:classes}
\resizebox{\columnwidth}{!}{%
\begin{tabular}{|l|l|}
\hline
\textbf{Class} & \textbf{Description}                                    \\ \hline
DataCenter                                & Models the physical facility housing HPC infrastructure.                      \\ \hline
HPCSystem                                 & Refers to an entire HPC system composed of multiple racks, compute nodes and other HPC infrastructure. 
\\ \hline
User                                      & Represents an individual user of the HPC system who submits workloads or jobs for execution on the HPC system.
\\ \hline
Job                                       & Describes a computational task submitted to the HPC system.                   \\ \hline
JobMetric                                 & Captures performance or requested resources metrics related to a job.         \\ \hline
Rack                                      & Models a rack structure used to hold compute nodes.                           \\ \hline
ComputeNode                               & Represents an individual computational node in an HPC system.                 \\ \hline
Position                                  & Defines the physical location of the compute node in three dimensions.        \\ \hline
Plugin                                    & Represents an extensible software component or module used within the system. \\ \hline
Sensor                                    & Denotes a hardware or virtual device that collects monitoring data.           \\ \hline
SensorReading                             & Represents the monitored data output from a sensor at a point in time.        \\ \hline
Time                                      & Encodes temporal information for time modeling in the proposed ontology.      \\ \hline
\end{tabular}%
}
\end{table}

% \subsection{Object Properties}
% \label{sec:obj_properties}

% Table~\ref{tab:obj_prop} presents 23 object properties in the proposed ODA ontology, which describe relationships between key entities in the HPC domain. The table shows the domain and range for each property, where the domain specifies the class from which the property originates, and the range defines the class to which the property applies. These properties link the classes listed in Table~\ref{tab:classes}, helping to represent the relationships within HPC infrastructures. Complementing these, Table~\ref{tab:data_prop} lists the data properties, which define the attributes of individual classes. Similar to object properties, each data property includes a domain and range; however, the range here corresponds to the datatype of the property value. These datatypes conform to the XML Schema Definition (XSD) standards, including types such as xsd:integer for integers, xsd:string for text, and xsd:dateTime for date-time values.

Table~\ref{tab:obj_prop} lists 23 object properties defined in the proposed ODA ontology, describing how key entities in the HPC system are related. Each property includes a domain, representing the source class, and a range, representing the target class. These properties establish semantic links between the classes listed in Table~\ref{tab:classes}. Complementing these, Table~\ref{tab:data_prop} lists the 25 data properties, which define attributes of individual classes. Like object properties, each data property includes a domain and range; however, the range here refers to the property's datatype, following XML Schema Definition (XSD) standards—such as xsd:integer for integers, xsd:string for text, and xsd:dateTime for dateTime values.

% Table~\ref{tab:properties}(b) lists the data properties in the proposed ODA ontology, which define attributes of each class of the ontology listed in the table~\ref{tab:classes}. Each property includes its associated domain (the entity to which it applies), its range (the type of value it holds), and a description. The ranges are specified using XML Schema Data Types (xsd), which are standard data types used to define the format and constraints of data withn an RDF ontology. For example, xsd:integer represents an integer value, xsd:string represents text, and xsd:dateTime represents a date and time.

% Table~\ref{tab:properties}(a) presents 23 object properties in the proposed ODA ontology, which define relationships between key entities in the HPC domain. Each property includes a domain (the class from which the property originates) and a range (the class to which the property points), enabling structured representation of interactions among the classes listed in Table~\ref{tab:classes}. Complementing these, Table~\ref{tab:properties}(b) lists the data properties, which specify the attributes of individual classes within the ontology. Like object properties, each data property includes a domain and a range; however, in this case, the range refers to the datatype of the property value. These datatypes follow the XML Schema Definition (XSD), such as xsd:integer for integer values, xsd:string for text, and xsd:dateTime for date-time values.

\vspace{-20pt}

\begin{table}[h]
\centering
\caption{Object properties defining relationships between classes within the proposed ontology.}
\label{tab:obj_prop}
\resizebox{\columnwidth}{!}{%
\begin{tabular}{|l|l|l|l|}
\hline
\textbf{Property} & \textbf{Domain} & \textbf{Range} & \textbf{Description} \\ \hline
hasHPCSystem    & DataCenter    & HPCSystem     & Associates a data center with the HPC systems it contains.    \\ \hline
isHPCSystemOf   & HPCSystem     & DataCenter    & Inverse of hasHPCSystem.                                      \\ \hline
hasUser         & HPCSystem     & User          & Associates a user with a particular HPC system.               \\ \hline
isUserOf        & User          & HPCSystem     & Inverse of hasUser.                                           \\ \hline
hasRack         & HPCSystem     & Rack          & Associates an HPC system with its constituent racks.          \\ \hline
isRackOf        & Rack          & HPCSystem     & Inverse of hasRack.                                           \\ \hline
hasComputeNode  & Rack          & ComputeNode   & Indicates that a rack contains a compute nodes.               \\ \hline
isComputeNodeOf & ComputeNode   & Rack          & Inverse of hasComputeNode.                                    \\ \hline
hasPosition     & ComputeNode   & Position      & Indicates the physical position of a compute node.            \\ \hline
isPositionOf    & Position      & ComputeNode   & Inverse of hasPosition.                                       \\ \hline
isJobOf         & Job           & User          & Indicates the user who submitted a job.                       \\ \hline
submitsJob      & User          & Job           & Inverse of isJobOf.                                           \\ \hline
usesComputeNode & Job           & ComputeNode   & Links a job to the compute node(s) it utilizes.               \\ \hline
hasJobStartTime & Job           & Time          & Specifies the start time of a job.                            \\ \hline
hasJobEndTime   & Job           & Time          & Specifies the end time of a job.                              \\ \hline
hasJobMetric    & Job           & JobMetric     & Links a job to its performance metrics and requested resources.              \\ \hline
hasPlugin       & ComputeNode   & Plugin        & Indicates plugins or modules installed on a compute node.     \\ \hline
hasReading      & Sensor        & SensorReading & Links a sensor to its recorded readings.                      \\ \hline
hasSensor       & ComputeNode   & Sensor        & Indicates sensors installed on a compute node.                \\ \hline
isSensorOf      & Sensor        & ComputeNode   & Inverse of hasSensor.                                         \\ \hline
hasTimestamp    & SensorReading & Time          & Specifies the time when a sensor reading was recorded.        \\ \hline
includesSensor  & Plugin        & Sensor        & Indicates sensors included as part of a plugin.               \\ \hline
isPartOfPlugin  & Sensor        & Plugin        & Inverse of includesSensor.                                    \\ \hline
\end{tabular}%
}
\end{table}

% Table~\ref{tab:data_prop} lists the data properties, which specify the attributes of individual classes within the ontology. Like object properties, each data property includes a domain and a range; however, in this case, the range refers to the datatype of the property value. These datatypes follow the XML Schema Definition (XSD), such as xsd:integer for integer values, xsd:string for text, and xsd:dateTime for date-time values.

\begin{table}[h]
\centering
\caption{Data properties of the classes in the proposed ontology.}
\label{tab:data_prop}
\resizebox{\columnwidth}{!}{%
\begin{tabular}{|l|l|l|l|}
\hline
\textbf{Property} & \textbf{Domain} & \textbf{Range} & \textbf{Description}                                                   \\ \hline
dcId          & DataCenter    & xsd:integer  & The identifier for a data center.                                              \\ \hline
dcName        & DataCenter    & xsd:string   & The name of the data center.                                                   \\ \hline
location      & DataCenter    & xsd:string   & The physical location of the data center.                                      \\ \hline
systemId      & HPCSystem     & xsd:integer  & The identifier for a specific HPC system.                                      \\ \hline
systemName    & HPCSystem     & xsd:string   & The name of the HPC system.                                                    \\ \hline
userId        & User          & xsd:integer  & The identifier for a user in the system.                                       \\ \hline
userName      & User          & xsd:string   & The name of a user in the HPC system.                                          \\ \hline
rackId        & Rack          & xsd:integer  & A unique identifier for a rack within the HPC system.                          \\ \hline
computeNodeId & ComputeNode   & xsd:integer  & A unique identifier for a compute node.                                        \\ \hline
posX          & Position      & xsd:integer  & The X-coordinate of a compute node’s position within a rack         \\ \hline
posY          & Position      & xsd:integer  & The Y-coordinate of a compute node’s position within a rack        \\ \hline
posZ          & Position      & xsd:integer  & The Z-coordinate of a compute node’s position within a rack        \\ \hline
pluginName    & Plugin        & xsd:string   & The name of a plugin installed on a compute node.                              \\ \hline
jobId         & Job           & xsd:integer  & A unique identifier for a job.                                                 \\ \hline
jobName       & Job           & xsd:string   & The name given to the job.                                                     \\ \hline
groupId       & Job           & xsd:integer  & The identifier for the group of the user who submitted the job.                                       \\ \hline
exitCode      & Job           & xsd:integer  & The exit code of a job upon completion, indicating success or failure.         \\ \hline
jobDuration   & Job           & xsd:duration   & The duration of job in the ISO 8601 duration format.      \\ \hline
metricName    & JobMetric     & xsd:string   & The name of the metric associated with a job’s performance or resource usage.  \\ \hline
metricValue   & JobMetric     & xsd:float    & The value of the performance or resource usage metric for a job.               \\ \hline
sensorName    & Sensor        & xsd:string   & The name of a sensor installed in a compute node.                   \\ \hline
sensorType    & Sensor        & xsd:string   & The type or category of the sensor (e.g., temperature, power).           \\ \hline
sensorUnit    & Sensor        & xsd:string   & The unit of measurement for the sensor’s data (e.g., °C, watts).               \\ \hline
timestamp     & Time          & xsd:dateTime & The timestamp associated with an event (e.g., job submission) or sensor reading.                      \\ \hline
value         & SensorReading & xsd:double   & The value of a sensor reading \\ \hline
\end{tabular}%
}
\end{table}

\section{Experimental Results}
\label{sec:experimental_results}

This section evaluates the proposed Unified ODA Ontology. We assess its effectiveness in unifying ODA telemetry from the Marconi100 (M100) \cite{m100nature} and Fugaku's F-DATA \cite{antici2024fdata}, and evaluate the proposed optimizations to reduce the storage overhead of the generated KG using the proposed ontology, compared to the previous implementation in \cite{khanExaQueryProvingData2024}.

\subsection{Experimental setting}
\label{sec:exp_setting}
\sloppy
The proposed ontology adopts the prefix \texttt{hpc} with the namespace URI \texttt{http://ontology.hpc.org/} as a placeholder. This URI is not currently resolvable or formally registered but serves to uniquely identify entities within the generated KG for experimental use. The ontology was created using Protégé, an open-source ontology editor, and is published in Turtle (TTL) format alongside OWLDoc documentation in the project’s Git repository(\url{https://gitlab.com/ecs-lab/unified-oda-ontology}). The ontology was designed to address a defined set of competency questions, which are subsequently used to evaluate its adequacy. The complete list of competency questions is presented in Table~\ref{tab:competency_questions}.

\begin{table}[ht]
\centering
\caption{Competency questions used to validate the proposed Unified ODA Ontology.}
\label{tab:competency_questions}
\resizebox{\columnwidth}{!}{%
\begin{tabular}{|l|l|l|}
\hline
\textbf{ID} & \textbf{Category} & \textbf{Competency Question}                                                 \\ \hline
C1.1        & System \&         & What HPC systems are present in data center Y?                               \\ \cline{1-1} \cline{3-3} 
C1.2        & Infrastructure    & What is the position of all the compute nodes in data center Y?              \\ \cline{1-1} \cline{3-3} 
C1.3        & Topology          & Which compute nodes are present in rack X of the Y system?                   \\ \cline{1-1} \cline{3-3} 
C1.4        &                   & Which sensors are installed on compute node N?                               \\ \cline{1-1} \cline{3-3} 
C1.5        &                   & Which compute nodes are in close spatial proximity to compute node N?        \\ \cline{1-1} \cline{3-3} 
C1.6        &                   & Which compute nodes are near to the spatial position (X,Y,Z)?                \\ \hline
C2.1 & Sensor \& & What is the max/min/avg recorded metric X for compute node N from T1 to T2?                                                     \\ \cline{1-1} \cline{3-3} 
C2.2        & Environmental     & What is the max/min/avg recorded metric X for rack R today?                  \\ \cline{1-1} \cline{3-3} 
C2.3        & Monitoring        & What is the max/min/avg recorded metric X for Job J?                         \\ \cline{1-1} \cline{3-3} 
C2.4        &                   & Are there any sensor readings above a certain threshold in the last hour?    \\ \cline{1-1} \cline{3-3} 
C2.5 &           & Can you tell me the list of compute nodes that had temperatures higher than normal last week?                                   \\ \cline{1-1} \cline{3-3} 
C2.6        &                   & Please provide me the compute node N average temperature during May 2025.    \\ \cline{1-1} \cline{3-3} 
C2.7        &                   & What is the average CPU utilization per compute node over the last 24 hours? \\ \cline{1-1} \cline{3-3} 
C2.8        &                   & What was the power consumption of compute node N at time T?                  \\ \hline
C3.1        & Job Metrics \&    & What is the power and energy consumption of job J?                           \\ \cline{1-1} \cline{3-3} 
C3.2        & Execution         & Which nodes did job J run on?                                                \\ \cline{1-1} \cline{3-3} 
C3.3        & Analysis          & Which jobs were running during time interval T1 to T2?                       \\ \cline{1-1} \cline{3-3} 
C3.4 &           & Which jobs are large scale? (i.e., they occupy a significant portion of the system's compute nodes)                             \\ \cline{1-1} \cline{3-3} 
C3.5        &                   & What was the duration of job J?                                              \\ \cline{1-1} \cline{3-3} 
C3.6        &                   & Which jobs failed due to insufficient walltime?                              \\ \cline{1-1} \cline{3-3} 
C3.7        &                   & Which jobs are memory/compute bound?                                         \\ \cline{1-1} \cline{3-3} 
C3.8 &           & Which jobs have incorrect frequency selection? (i.e. frequency at 2200 and memory bound or frequency at 2000 and compute bound) \\ \cline{1-1} \cline{3-3} 
C3.9        &                   & What is the utilization of CPUs and GPUs during the jobs executed by user U? \\ \hline
C4.1        & User \&           & Which user submitted job J?                                                  \\ \cline{1-1} \cline{3-3} 
C4.2        & Group             & Which users belong to group G?                                               \\ \cline{1-1} \cline{3-3} 
C4.3        & Activity          & Which users submitted jobs during period T?                                  \\ \cline{1-1} \cline{3-3} 
C4.4        &                   & How many jobs has user U submitted this week?                                \\ \cline{1-1} \cline{3-3} 
C4.5        &                   & Which users have submitted the most memory bound jobs?                       \\ \cline{1-1} \cline{3-3} 
C4.6        &                   & Which users submitted jobs in Data Center Y?                                 \\ \hline
C5.1        & Scheduling \&     & What is the average waiting time of the jobs in the queue?                   \\ \cline{1-1} \cline{3-3} 
C5.2        & Queue Analysis    & Which jobs waited more than X hours in the queue?                            \\ \hline
C6.1        & Cross-System \&   & Which data centers have the highest job submission volumes?                  \\ \cline{1-1} \cline{3-3} 
C6.2        & Comparative       & What job metrics are present in one HPC system but not others?               \\ \cline{1-1} \cline{3-3} 
C6.3        & Analytics         & How does average job execution time vary across HPC systems?                 \\ \cline{1-1} \cline{3-3} 
C6.4        & \multirow{2}{*}{} & Which system is more energy-efficient per job executed?                      \\ \cline{1-1} \cline{3-3} 
C6.5        &                   & When is the low power consumption period for system Y?                       \\ \hline
\end{tabular}%
}
\end{table}

\subsection{Validation of the proposed Unified ODA Ontology}
\label{sec:validation_ontology}

The proposed ontology was validated by mapping its relevant classes and object properties to the competency questions defined in each category (see Table~\ref{tab:competency_questions}). For each category, we identified the key classes and assessed whether their relationships provided sufficient information to answer all questions accurately. For \textbf{System \& Infrastructure Topology} (C1.1–C1.6), the important classes are \texttt{DataCenter}, \texttt{HPCSystem}, \texttt{Rack}, \texttt{ComputeNode}, and \texttt{Position}. Together with properties like \texttt{hasComputeNode}, \texttt{hasRack}, and \texttt{hasPosition}, they support all questions in this category. In the \textbf{Sensor \& Environmental Monitoring} (C2.1--C2.8) category, the classes \texttt{Sensor} (representing physical sensors), \texttt{SensorReading} (capturing individual measurements), and \texttt{Time} (supporting the temporal aspects of the questions)—together with properties such as \texttt{hasReading} and \texttt{hasTimestamp}—collectively support all questions in this category. For the \textbf{Job Metrics \& Execution Analysis} (C3.1-C3.9), the \texttt{Job} class---connected to the \texttt{JobMetric} class, which holds job-specific metrics, and to the \texttt{ComputeNode} class, where the job was executed---together with their associated properties, provides sufficient support to answer the questions in this category. In \textbf{User \& Group Activity} (C4.1–C4.6), the \texttt{User} class identifies the HPC system users, and the \texttt{Job} class connects these users to their submitted jobs, and \texttt{HPCSystem} associates users with specific systems. Together, they support all user-related queries. For the \textbf{Scheduling \& Queue Analysis} (C5.1--C5.2), the \texttt{Job} and \texttt{JobMetric} classes, along with their associated properties, provide sufficient support to answer questions in this category. Finally, in the \textbf{Cross-System \& Comparative Analytics} (C6.1–C6.5) category, the key classes are \texttt{DataCenter} and \texttt{HPCSystem}. In combination with other classes—from \texttt{User} to \texttt{Job} for job-specific analysis, and the \texttt{Rack} and \texttt{ComputeNode} classes, with their associated properties linking to specific sensor readings within each HPC system, allow for effective cross-system analysis. Listing~\ref{lst:C6.3} presents the SPARQL query corresponding to competency question C6.3 from the \textbf{Cross-System \& Comparative Analytics} category, as defined in Table~\ref{tab:competency_questions}. This query illustrates how the ontology facilitates cross-system comparisons. The complete set of SPARQL queries for all competency questions is available in the project’s Git repository.

\begin{lstlisting}[language=SPARQL, caption={SPARQL query for the competency question C6.3}, label={lst:C6.3}, basicstyle=\ttfamily\scriptsize, breaklines=true, frame=single, columns=flexible, numbers=left, float=h]
PREFIX hpc: <http://ontology.hpc.org/>
PREFIX xsd: <http://www.w3.org/2001/XMLSchema#>

SELECT ?hpcSystem ?systemName (AVG(?execTimeSeconds) AS ?avgExecutionTimeSeconds)
WHERE {
  ?hpcSystem a hpc:HPCSystem ;
             hpc:systemName ?systemName ;
             hpc:hasRack ?rack .
  ?rack hpc:hasComputeNode ?computeNode .
  ?job hpc:usesComputeNode ?computeNode ;
       hpc:hasJobStartTime ?startTime ;
       hpc:hasJobEndTime ?endTime .
  ?startTime hpc:timestamp ?startTimestamp .
  ?endTime hpc:timestamp ?endTimestamp .

  BIND(
    ( xsd:dateTime(?endTimestamp) - xsd:dateTime(?startTimestamp) ) * 86400 AS ?execTimeSeconds
  )
}
GROUP BY ?hpcSystem ?systemName
ORDER BY ?hpcSystem
\end{lstlisting}

\subsection{Storage size optimizations}
\label{sec:storage_size}

\vspace{-20pt}
\begin{table}[ht]
\centering
\caption{Comparison of storage size, triple count, and node count for KGs generated using the ODA ontology \cite{khanExaQueryProvingData2024} and the proposed Unified ODA Ontology, with and without blank nodes (bNodes) for sensor readings.}
\label{tab:size_comparisons}
\resizebox{\columnwidth}{!}{%
\begin{tabular}{|l|r|r|r|}
\hline
\textbf{Version}                          & \textbf{\# Triples} & \textbf{\# Nodes} & \textbf{Size [MiB]} \\ \hline
ODA ontology \cite{khanExaQueryProvingData2024} & 25,375,684          & 4,234,662         & 1074.89             \\ \hline
Proposed Unified Ontology                 & 16,917,120          & 4,234,658         & 657.36              \\ \hline
Proposed Unified Ontology {[}bNodes for Sensor Reading{]} & 16,917,120 & 4,234,658 & 481.00 \\ \hline
\end{tabular}%
}
\end{table}

\vspace{-10pt}

Table~\ref{tab:size_comparisons} presents the results for the comparison of the storage requirements of KG constructed using the ODA ontology implementation from \cite{khanExaQueryProvingData2024} with those constructed using the proposed unified ontology under two configurations: (1) assigning URIs to each sensor reading, and (2) representing sensor readings as blank nodes (i.e., anonymous RDF resources without globally unique URIs). Blank nodes can help reduce storage size by eliminating the need to generate and store long URI strings for entities that do not require external reference, thereby decreasing storage overhead. For this analysis, we used one day of sensor data from the M100 dataset \cite{m100nature}, specifically the "total\_power" metric collected by the IPMI plugin on February 1st, 2022. For reference, the same data requires only 2.77 MiB when stored in a NoSQL DB, highlighting the additional overhead introduced by semantic modeling and RDF serialization. The key metrics evaluated include the number of RDF triples, the number of RDF nodes, and the storage size in MiB. The KG constructed using the ODA ontology in \cite{khanExaQueryProvingData2024} consists of approximately 25.4 million triples and over 4.2 million nodes, requiring 1,074.89 MiB of storage. In contrast, the KG generated using the proposed unified ontology with URIs for sensor readings uses about 16.9 million triples and the same number of nodes, reducing the storage size to 657.36 MiB—an approximate 38.84\% reduction. This improvement results from design optimizations introduced during the ontology design process (see Section~\ref{sec:ontology_design_process}), including removal of DataRecord redundancies, introduction of the Time class, and moving unit information to the Sensor class. Furthermore, using the second configuration—modeling sensor readings as blank nodes—reduces the size to 481.00 MiB, achieving a further reduction of 26.82\%.

\section{Discussion}
\label{sec:discussion}

The evaluation primarily demonstrates storage size reductions achieved through ontology design optimizations. Although limited to one day of sensor data for a single telemetry metric (“total\_power”), the results show significant improvements in storage efficiency compared to previous work.

Storage requirements for ODA telemetry data scale linearly with both sampling frequency and measurement duration, as each data point generates a predictable number of RDF triples. For example, at a 20s sampling interval, the “total\_power” metric in the M100 dataset \cite{m100nature} requires approximately 2.8 MiB per day, closely matching the 2.77 MiB observed in this single-day evaluation. This consistency confirms proportional RDF graph growth, enabling reliable projection of storage needs over longer periods. Consequently, storing this metric for February 2022 would require about 29.39 GiB using the previous approach \cite{khanExaQueryProvingData2024}, while the proposed unified ontology reduces this to 17.97 GiB without blank nodes or 13.15 GiB with blank nodes. This linear scaling can be similarly extended to other metrics and sensors in the dataset.

Despite these significant reductions, the generated KG sizes remain larger than equivalent data stored in NoSQL (approx 238× without blank nodes, 174× with blank nodes). This raises the question of the practical justification for semantic modeling given the storage overhead. The answer lies in KGs’ advantages for query flexibility and usability. As shown in \cite{khanExaQueryProvingData2024}, KGs simplify writing complex ODA queries in SPARQL compared to commonly used NoSQL/SQLite queries \cite{ODA}. Furthermore, recent work \cite{khan2025exasage} demonstrates that large language models (LLMs) automate SPARQL query generation with far greater accuracy—92\% versus roughly 25\% for equivalent NoSQL/SQLite queries. Considering these benefits, improving storage efficiency is important to maintain the practicality of this powerful and user-friendly approach at scale despite increased storage demands.

\section{Conclusion}
\label{sec:conclusion}

% In this paper, we proposed a Unified ODA Ontology for data center and HPC systems. We detailed the ontology design process and described all associated properties. The proposed ontology was validated using a set of 36 competency questions, demonstrating that its classes and properties are sufficient to answer them. We also compared the storage size requirements of a KG built using the implementation from \cite{khanExaQueryProvingData2024} with that of the proposed Unified ODA Ontology, evaluated under two deployment configurations: with and without blank nodes for the sensor reading class. The Unified ODA Ontology achieved a 38.84\% reduction in storage size when using URIs for sensor readings, and a further 26.82\% reduction when using blank nodes. While still larger than the NoSQL equivalent (2.77 MiB), the proposed Unified ODA Ontology provides a semantically rich foundation for advanced querying. The achieved storage reductions represent a key step towards scalable semantic modeling of ODA telemetry data. The proposed ontology successfully unifies the two largest publicly available ODA datasets of M100\cite{m100nature} and F-DATA\cite{antici2024fdata}.

In this paper, we propose a Unified ODA Ontology for data centers. We describe its design process and detail its classes and properties. The ontology was validated using 36 competency questions, demonstrating its suitability for answering them. It successfully unifies the two largest publicly available ODA datasets: (1) M100 \cite{m100nature} and (2) F-DATA \cite{antici2024fdata}. We compared the storage footprint of a KG built with the proposed ontology against the implementation in \cite{khanExaQueryProvingData2024}, under two configurations: (1) using URIs and (2) using blank nodes for sensor readings. Our ontology achieved a 38.84\% reduction in storage with URIs, and a further 26.82\% reduction with blank nodes. While the storage footprint remains larger than the NoSQL equivalent (2.77 MiB), the ontology’s semantically rich structure enables advanced querying, and the achieved reductions represent a significant step toward scalable semantic modeling of ODA telemetry data. Future work will expand the evaluation to more datasets and metrics, include query performance analysis, and implement further storage optimizations. We also plan to provide a persistent, resolvable ontology IRI supported by a comprehensive strategy for long-term hosting and governance to facilitate reuse and adoption.

%While still larger than the NoSQL equivalent (2.77 MiB), the proposed Unified ODA Ontology offers a semantically rich foundation for advanced querying, integration, and reasoning. The achieved storage reductions mark a key step toward efficient, scalable semantic modeling, while also enabling seamless integration with LLMs for automated query generation and code synthesis in HPC and data center environments.

\section*{Acknowledgments}

% This research was partly supported by the HE EU DECICE project (g.a. 101092582), the HE EU Graph-Massivizer project (g.a. 101093202), and \\ SPOKE 1: Future HPC \& Big Data by PNRR. We also thank CINECA for their collaboration and for providing access to their computing resources.

This research was supported by EuroHPC JU SEANERGYS (g.a. 101177590), HE EU DECICE (g.a.\ 101092582), HE EU Graph-Massivizer (g.a.\ 101093202), and SPOKE 1: Future HPC \& Big Data (PNRR). We thank Francesco Antici (who collected the F-DATA dataset from Fugaku) and CINECA for their collaboration in formulating the competency questions as HPC experts.
 \bibliographystyle{splncs04}
 \bibliography{references}




% LNCS Homepage, \url{http://www.springer.com/lncs}, last accessed 2023/10/25
% \end{thebibliography}
\end{document}